# Surfaces and orientations: much to fret about?

Ivan Rasnik[1], Sean A. McKinney[1] and Taekjip Ha[1,2]
[1]Physics department and [2]Center for biophysics and computational biology, University of Illinois, Urbana-Champaign, Urbana, IL 61801, USA

*Abstract: Single molecule FRET (fluorescence resonance energy transfer) is a powerful technique for detecting real-time conformational changes and molecular interactions during biological reactions. In this review, we examine different techniques of extending observation times via immobilization and illustrate how useful biological information can be obtained from single molecule FRET time trajectories with or without absolute distance information.*

Single-molecule fluorescence imaging is a powerful tool for probing biological events directly without the temporal and population averaging of conventional ensemble studies [reviewed in 1, 2, 3] One particularly popular incarnation is single molecule FRET (fluorescence resonance energy transfer). FRET is a spectroscopic technique for measuring distances in the 30-80 Å range [4, 5]. Excitation energy of the donor is transferred to the acceptor via an induced-dipole, induced-dipole interaction. The efficiency of energy transfer, $E$, is given by $[1+(R/R_0)^6]^{-1}$ where $R$ is the distance between the donor and acceptor. $R_0$ is the distance at which 50% of the energy is transferred and is a function of the spectral overlap of the donor emission and acceptor absorption, refractive index of the medium, quantum yield of the donor and a factor $\kappa^2$ that depends on the relative orientation in space between the transition dipoles for donor and acceptor. A small change in distance between the two sites of a biological molecule where donor and acceptor are attached can result in a sizeable change in $E$. Therefore, structural changes of biological molecules or relative motion between two different molecules can be detected via FRET changes [6]. The efficiency of energy transfer can be measured by determining the donor lifetime or its fluorescence intensity in presence and absence of the acceptor. Alternatively, as we have done in the works presented here, it can be obtained by calculating $I_D/(I_D + I_A)$, the ratio between the donor fluorescence intensity and the total intensity (donor plus acceptor). For accurate determinations the previous expression must be corrected for leakage of donor fluorescence in the acceptor channel and vice versa, corrections that can be easily determined with samples presenting donor-only or acceptor only signals.

Often, conformational changes are very difficult to synchronize or occur too infrequently to detect using ensemble FRET. Single molecule FRET [reviews in 1, 2,

7-10] opens up new opportunities to probe the real time structural changes of biological molecules. In addition, single molecule FRET readily determines not just average conformations, but also the distribution of distances. Single molecule FRET is also relatively insensitive to incomplete labeling of host molecule with donor and acceptor. Donor-only species simply show up as FRET=0 species while acceptor-only species are excited only very weakly if the probes are selected with a large spectral separation. Since the first demonstration of single molecule FRET [11], there have been a number of experiments geared toward biological applications. A partial list includes the oligomerization of membrane proteins in a living cell [12, 13], protein folding [14-18], protein conformational changes [19-25], RNA folding and catalysis [26-36], DNA structural dynamics [34, 37-41], and DNA-protein interactions [42-45].

As of now there are two primary methods of performing single molecule FRET measurements: in solution and on the surface. Although solution experiments are often used [14, 16, 17, 37, 45] and completely alleviate concerns about surface interactions, their diffusion limited temporal window leaves them inadequate for investigation of slower (>10ms) phenomena. Surface immobilization provides extended observation window limited only by photobleaching, but requires careful attention because of potential artifacts induced by surface interactions. Here we review recent works using the various types of immobilization schemes. We also review attempts to extend the power of single molecule FRET in resolving potential ambiguities and in obtaining quantitative distance information.

## 1. Surface effects

Tracking asynchronous time evolution of single biological molecules provides unique insight into detailed reaction kinetics and pathways. This requires prolonged observation periods in biologically relevant environments. Such measurements are frequently made on macromolecules that are tethered on a glass surface. Several different immobilization methods have been employed, including specific single-point attachment on glass [46] or polymer coated surfaces [43, 44, 47], and confinement inside porous polyacrylamide [48] and agarose [49] gel matrices and inside phospholipid vesicles [18, 50, 51]. Here we will discuss many of these approaches, illustrated in Fig. 1.

*BSA-biotin surface:* For DNA or RNA only studies, quartz or glass surfaces first coated with biotinylated BSA and then streptavidin are ideal. Binding of nucleic acids modified with biotin to this surface is highly specific; no binding is observed if either biotin or streptavidin is omitted. All previous studies have shown that average behaviors of single DNA or RNA molecules are identical to their bulk solution behaviors [26, 27, 32, 34, 40]. This is very likely because all three surface constituents (quartz, BSA and streptavidin) are negatively charged in neutral pH, repelling the likewise negatively charged nucleic acids away from the surface except at the biotinylated attachment point. However in some cases, such as in the folding kinetics of the hairpin ribozyme, very significant heterogeneity (two to three orders of magnitude) was observed from molecule to molecule in folding and unfolding rates [32, 34]. It remained to be shown whether this heterogeneity was the result of molecular interactions of the RNA with the surface, or intrinsic complex behavior of macromolecules.

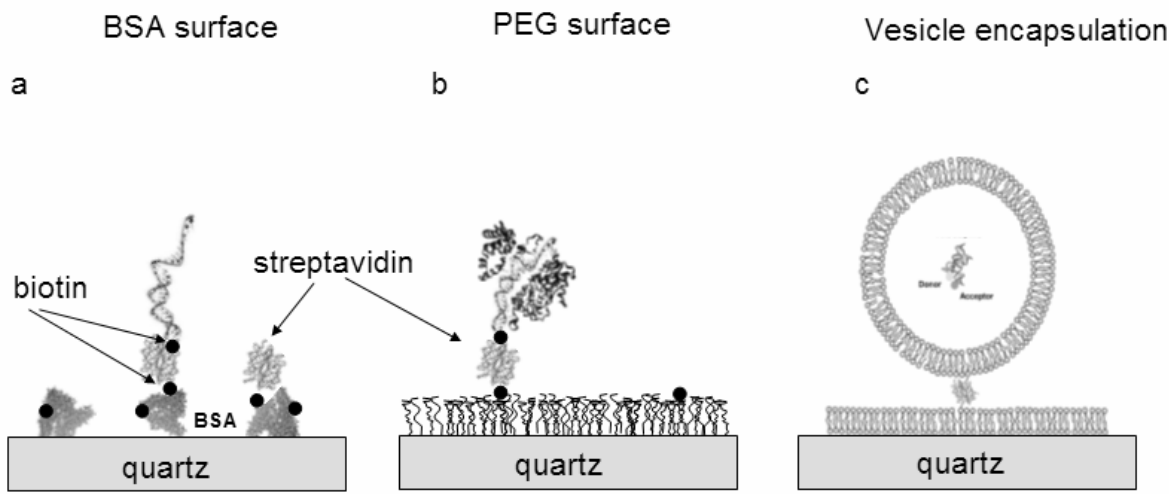

**Figure 1** Three different immobilization techniques. **a.** Streptavidin molecules bind to biotin-BSA molecules non-specifically attached to a glass or quartz slide. The remaining free pockets of the streptavidin can bind biotin-modified molecules providing a simple and reliable immobilization protocol for oligo nucleotides, but not for proteins. **b.** High density of amino groups are introduced on a quartz surface by aminosilanization. The amino groups are then reacted with NHS-ester modified PEG, creating a polymer brush covalently attached to the surface. Using a small fraction of biotin-PEG-NHS ester allows binding of biotin-modified molecules. **c.** Biomolecules are encapsulated in small unilamelar vesicles that are then immobilized through biotin-streptavidin binding to a supported lipid bilayer on the surface of a quartz slide. The biomolecule is free to diffuse inside the volume of the vesicle.

*Vesicle encapsulation:* To see if the surface was causing heterogeneities, Okumus *et al* encapsulated the hairpin ribozyme in 100 nm-diameter vesicles and tethered the vesicles to a supported bilayer, following the general protocol developed by Haran and co-workers for protein folding studies [18, 50]. This allowed long observation times while avoiding direct surface attachment of the RNA. RNA molecules did not bind to the membrane stably even at 100 nM concentration, meaning non-specific binding of RNA molecules to membranes, if at all, is short lived. As Fig. 2 shows, the highly heterogeneous folding kinetics of the hairpin ribozyme are almost identically reproduced when the molecule is encapsulated [51]. Since these heterogeneities are long lasting (a molecule displays its own folding/unfolding rates for minutes) and the above mentioned control experiment showed that there is no stable binding of RNA to the membrane, we can now safely exclude surface effects as the source of heterogeneity. These studies show that vesicle encapsulation is an excellent means of immobilizing RNA molecule (likely true for DNA as well) for single-molecule observation without direct surface tethering. Unfortunately phospholipid membranes are impermeable to small molecules such as ATP and ions, limiting the utility of vesicle encapsulation. In contrast, direct surface tethering of biomolecules allows the triggering of biochemical reaction by flow (for instance $Mg^{2+}$ addition for RNA folding and catalysis [27], and ATP addition for DNA unwinding by helicase [43]).

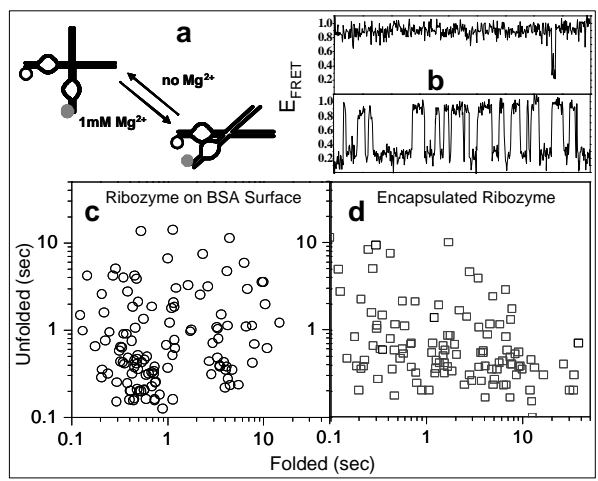

**Figure 2** Heterogeneous folding kinetics of single ribozymes are reproduced inside the vesicles. **a.** Hairpin ribozyme folding and unfolding can be followed by FRET between the donor (white sphere) and the acceptor (gray sphere). The folded state presents high FRET values while the unfolded state shows low FRET values. **b.** FRET time traces under the same buffer conditions reveal very different dynamic behavior for individual molecules. **c.** Plot of the folded state dwell time vs. unfolded state dwell time for molecules on a BSA surface show highly heterogeneous behavior. **d.** The same distribution is observed for molecules immobilized by vesicle encapsulation.

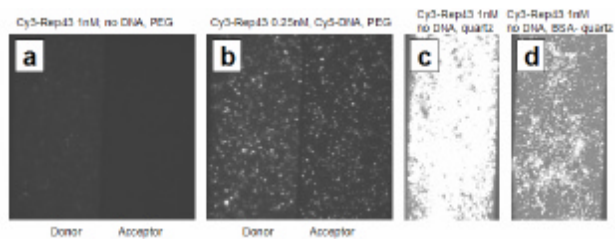

**Figure 3** PEG coating of surfaces prevent non-specific binding of proteins. a. As a control, a 1 nM solution of Cy3 labeled protein (*E. coli* Rep helicase) is applied to a sample in the absence of immobilized DNA. For a PEG coated surface the image on the CCD camera (an average over 10 frames) shows that binding to the surface is negligible. b. After immobilization of Cy5 labeled DNA molecules to the PEG coated surface via biotin-streptavidin individual spots can be observed on the CCD camera even at 0.25 nM concentration of Cy3 labeled Rep. The presence of spots in both the Cy3 and Cy5 channels shows that proteins are binding to the DNA giving rise to FRET. c. When the control experiment is performed on a naked quartz surface a large amount of non-specific binding is observed. d. While the use of BSA coating reduces the amount of non-specific binding to the surface, the image shows that a large fraction of the surface is still covered by labeled proteins indicating that PEG is essential for preventing non-specific binding of proteins.

*PEG surface:* For studies involving some proteins, particularly proteins with positively charged binding domains, BSA-coated surfaces are too adhesive [44]. Therefore, we developed a PEG (polyethylene glycol)-coated surface that reduces the protein adsorption to an undetectable level [43, 52, 53]. If a dense layer of PEG is formed on a quartz surface (we use amino-silane coating followed by conjugation with PEG modified with NHS-ester at one end), it forms a polymer brush that prevents protein adsorption to the underlying surface. A small fraction of PEG polymers that are end-modified by biotin are incorporated to facilitate the immobilization of biotinylated macromolecules. Rasnik *et al* showed that the PEG surface completely rejects the adsorption of fluorescently labeled *E. coli* Rep helicases ( an enzyme that catalyzes the unwinding of double stranded DNA), while allowing their specific binding to surface immobilized DNA (Fig. 3) [44]. In addition, relative changes in FRET between different labeling sites on the protein were exactly reproduced in single-molecule and bulk solution measurements. An earlier work showed that the unwinding rate by the Rep helicase in single-molecule measurements on PEG surfaces is identical to that measured in bulk solution [44]. Our preliminary studies indicate that the PEG surface is highly suitable for many other nucleic acids-protein interactions. Regular linear PEG molecules, however, interact strongly with unfolded proteins, presumably due to hydrophobic interactions which prevent refolding. Nienhaus and colleagues overcame this problem by using branched polymers called "star-PEG" which enabled the observation of reversible folding and unfolding reactions from single protein molecules [54].

## 2. Advances in single molecule FRET

*Measuring absolute distances:* While single molecule FRET is an excellent technique for measuring relative distance changes, it is not ideal for obtaining absolute distances because calculating an R value requires knowing $R_0$ with some precision, which even for a well characterized dye pair with a well known spectral overlap and solvent response still requires some knowledge of relative dye orientation. Fortunately, in most cases it is not necessary to find absolute distances from FRET data. For instance in certain experiments for RNA folding, the state that becomes more populated at higher $Mg^{2+}$ can be identified as the folded form [26, 32, 55]. The Holliday junction, a recombination intermediate comprised of four single strands entangled in a criss-cross, similarly folds and unfolds, but this time FRET data could be compared to gel mobility data to assign the two FRET states to the two different conformations of the junction [40]. One can then use FRET values to estimate distances and compare to the predicted distances based on known structural information to test if these assignments are reasonable. In most experiments, therefore, distance information plays only a supporting role.

Can we get reliable absolute distance information? First some other possible sources of non-distance related FRET changes must be ruled out. To test if measured FRET values are affected by the particular dyes chosen, we used different donors of similar spectral properties (TMR or Alexa568 instead of Cy3) and swapped Cy3 and Cy5, and moved them to different positions; the same results were obtained for helicase-DNA binding geometries [44] and Holliday junction dynamics [41] (Fig. 4). The results suggest that FRET gives a robust measure of relative distances regardless of how specific dyes may interact with the host molecule. The observed trend for FRET efficiency as a function of labeling position is independent of the donor-acceptor pair. While the absolute distances obtained for a particular combination of donor-acceptor pair may have small variations, depending on the specific value of $R_0$, the consistency of results allows ruling out any aberrant behavior.

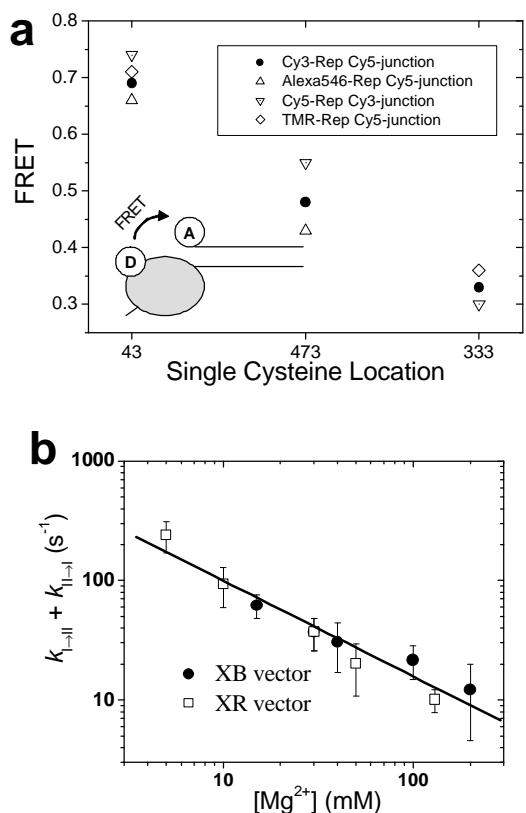

**Figure 4** Effects of Donor-Acceptor pair selection and dye's location on biomolecules for single molecule FRET experiments. a. The graph shows FRET efficiency values for three different labeling sites on the protein and for various donor-acceptor combinations for a protein-DNA interaction study. The position of the dye on the DNA was fixed at the junction of the partial duplex substrate used for the experiments. While there are small variations in FRET depending on the dyes chosen, the difference between labeling sites is much larger. b. Studies of conformational changes of a four way DNA junction were repeated by labeling different arms of the junction. The rate of transitions (sum of forward and backward rates) as a function of the magnesium concentration, based on dwell time analyses of single molecule trajectories, shows results that fall on a single line (a linear fit in the log-log scale) within error.

Sometimes FRET changes can not be assigned to known molecular conformations as in the case where single-molecule measurements reveal previously unknown conformations. Such FRET changes could potentially be due to changes in dye orientation without any change in the molecular conformation. The local environment may also affect the absolute value of $R_0$ through changes in the photophysical properties of the dyes or the effective refractive index of the medium.

One can address the issue of dyes orientation by carrying out simultaneous measurements of FRET and

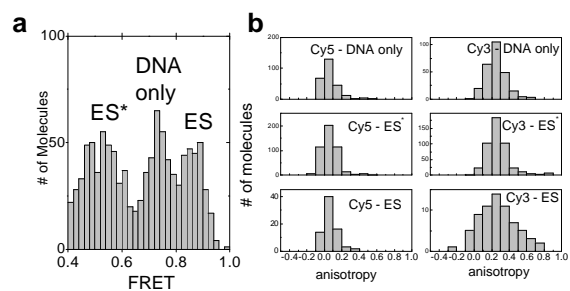

**Figure 5** Simultaneous determination of FRET and anisotropy from individual molecules. **a.** Partial duplex DNA was labeled with a Donor-Acceptor pair at the ends of the single-stranded tail to study binding of Rep helicase. Histograms of FRET efficiency values obtained from analyses of single molecule images show two distinct values at 0.5 (ES*) and 0.85 (ES) besides the peak corresponding to the DNA only (no protein bound) species. **b.** Using polarization optics for the excitation and the detection we split the image on the CCD camera based on color and polarization, which allows us to determine anisotropy values for each molecule. The histograms of anisotropy for molecules grouped by FRET efficiency values show no differences in anisotropy and are consistent with ensemble measurements. The broader distribution for Cy3-ES is fluorescence anisotropy from single molecules [45, 56]. If the orientational properties of the dye change substantially, this is likely to be detectable as a change in anisotropy. If there is no change in anisotropy, we can conclude that FRET changes are likely due to distance change. We have carried out this type of experiment for helicase/DNA binding (Fig. 5). Three different FRET states were observed corresponding to different conformations of single stranded DNA. By comparison with the FRET efficiency histograms obtained in absence of protein, the peaks at lower and higher FRET efficiency are assign ed to two different conformations of Enzyme-Substrates complexes of Rep bound to the DNA substrate (ES and ES* respectively). Fluorescent anisotropy histograms were obtained for each of the three sub-populations (Fig. 5). The data show that all three FRET sub-populations have identical anisotropy, indicating that they have different average distances.

Although these measurements clearly rule out the changes in FRET as a result of changes in dye orientations, in order to calculate exact distances one still needs to know the actual orientational term, $\kappa^2$. In almost all previous single molecule FRET studies that deduce distances, it was assumed that $\kappa^2=2/3$, not because this assumption is strictly valid, but for lack of further information. In fact, most fluorophores show relatively high fluorescence anisotropy and their orientations relative to the host molecules are not entirely randomized within their fluorescence lifetimes. Therefore, the best we can hope for is an approximation to a true distance. This process may be made more reliable via overly redundant experimentations. Following this avenue Rasnik *et al* performed single molecule FRET measurements on protein/DNA complexes, Rep helicase monomer/partial duplex DNA (Fig. 6). The goal was to determine the position of the DNA junction relative to the protein by measuring distances between the acceptor at the junction and the donor attached at one of the eight sites on the protein.

From a previously published crystal structure [57] it is known that Rep presents two very distinct conformations, corresponding to a swiveling motion of the 2B domain, one of the four structural domains of the monomer. Eight mutants of Rep protein were constructed, each containing only one cysteine to which the donor (Cy3) is conjugated.

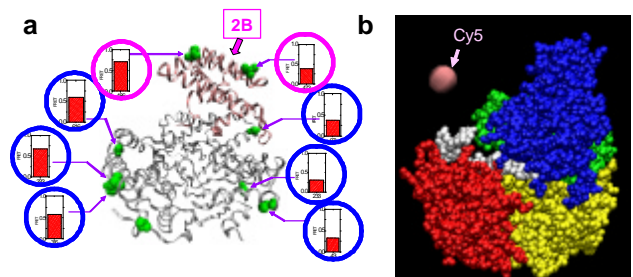

**Figure 6** Determination of absolute distances by multiple single molecule FRET experiments. a. We study binding of Rep monomer to a partial duplex DNA substrate. An acceptor molecule is present at the junction of the partial duplex and a donor molecule is positioned at different sites of the protein for each experiment through site-specific labeling to a single cysteine. From each experiment we obtain a FRET efficiency value, shown as bar graph, associated with the distance between the dye located at the protein and the dye on the DNA. Results from sites on the flexible 2B domain are highlighted with pink circles. b. With four FRET efficiency constraints it is possible to obtain a unique solution for the position of the acceptor molecule relative to the protein. Oversampling, in this case using six constraints, increases the accuracy of the process.

Since two residues are attached to the 2B domain six distance constraints from non-2B sites were used to deduce the acceptor location, assuming $\kappa^2=2/3$. This is an oversampling (or redundancy) because only four distance constraints are needed to locate the acceptor position uniquely. It is likely that potential site-specific effects on dye properties such as fluorescence anisotropy are averaged out via such oversampling. This process could be performed with a self-consistency of 5-10 Å for the acceptor localization and the junction location deduced this way is very similar to what was observed in the crystal structure of a complex between DNA and PcrA, a highly homologous helicase to Rep [58]. The remaining two distance constraints arising from two 2B sites were used to conclude that the 2B has the closed conformation, consistent with the crystal structure of PcrA bound to a partial duplex DNA. This example shows that the apparently crude approximation of $\kappa^2=2/3$ can still lead to useful information about biomolecular conformations. Variations in the other parameters that affect $R_0$ the can also be compensated by the oversampling method suggested above.

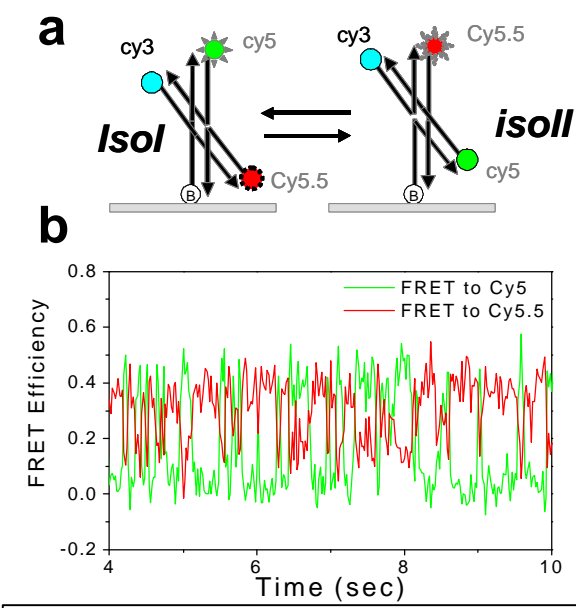

**Figure 7** Three color FRET experiments. **a.** A four way DNA junction is labeled with one donor (Cy3) and two acceptor molecules (Cy5 and Cy5.5) in different arms. The junction undergoes conformational changes between two species Iso I and Iso II. The dynamics of the transitions were previously studied by two colors FRET. **b.** A typical single molecule FRET trajectory showing FRET efficiency values for the Cy5 and Cy5.5 channel respectively. The perfectly anticorrelated behavior of the signals show a concerted movement of the arms of the junction that would be impossible to detect with other methods.

*Three-color FRET:* There are some cases in which regular single molecule FRET gives ambiguous information especially if the molecular system under study is complex. One can lift this ambiguity by measuring more than one distance at a time. For this purpose, Hohng *et al* developed three-color FRET using the DNA four-way (Holliday) junction as a model system [59]. Three fluorophores (Cy3, donor; Cy5, acceptor 1; Cy5.5, acceptor 2) are attached to three arms of the junction so that the donor would transfer energy alternatively to Cy5 or Cy5.5 upon conformer transitions (Fig. 7). With a judicious choice of fluorescent filters and careful correction for crosstalk between detection channels, they succeeded in measuring Cy5 and Cy5.5 alternatively lighting up as expected. The transition rates were identical to what were measured via two-color FRET [40, 41]. These anti-correlated fluctuations of FRET to Cy5 and to Cy5.5 were fully synchronized within 20 ms time resolution, demonstrating that the Cy5 arm is moving toward the Cy3 arm at the same time that the Cy5.5 arm is moving away from the Cy3 arm. While this was not a surprising result by itself, such a conclusion can not be drawn from regular two-color FRET studies. It is very likely that three-color FRET can help resolve potential ambiguities in a variety of systems. Another method of resolving some ambiguities in two-color single molecule FRET is the use of a separate laser to excite the acceptor directly so that the donor-only case can be distinguished from the cases with very low FRET [60].

## Conclusions
Single molecule FRET emerged as a potentially powerful technique to reveal dynamical information on biological systems. The distance dependence of FRET together with the power of single molecule analyses make the technique promising as a way to access information that could not be obtained by other methods. The works reviewed here show that experiments can be carried out in ways that minimally perturb the biological systems under study. At this point, different papers have shown independently that results, obtained with the different labeling and immobilization techniques available, are reliable and do not introduce artifacts if carried out properly. The possibility of using single molecule FRET as a nanoscopic ruler for absolute distances determination has also been shown, and this opens new possibilities in terms of structural analyses of biomolecular systems. Distance information obtained from a single configuration of donor-acceptor pair should not be considered absolutely reliable. However, it is possible via oversampling, measuring multiple donor-acceptor configurations, to obtain accurate models with reliable distance information. With well defined protocols and control experiments available, the limitations and potential of the technique well established, the next challenge resides in fully exploiting it to obtain previously unattainable data on interesting biological problems.

## Biosketches
Ivan Rasnik was born in Paysandú, Uruguay in 1967. After his undergrad studies in Chemistry in Montevideo, Uruguay, he received his PhD in Physics from Sao Paulo State University at Campinas, Brazil, in 2000. His research work was primarily on the optical properties of semiconductor heterostructures. He was a postdoctoral researcher at University of Pennsylvania in 2000, where he worked in quantum chemistry calculations of the optical properties of biomolecules. Since then he has been a postdoctoral researcher at University of Illinois, with his research focused on the development and use of single molecule fluorescence techniques to study biological problems.

Sean McKinney received a B.S. in physics and computer science from the University of Missouri at Rolla in 2001. In the summer of 2001 he joined Professor Taekjip Ha's laboratory in the University of Illinois at Urbana-Champaign in pursuit of his doctorate in physics. He is a recipient of 2003 National Science Foundation (NSF) graduate fellowship.

Taekjip Ha was born in Seoul, Korea in 1968. He obtained a B.S. in physics from Seoul National University in 1990, and a Ph.D in physics from University of California at Berkeley in 1996. After postdoctoral training in Lawrence Berkeley National Laboratory and Stanford University, he joined the University of Illinois, Urbana-Champaign in 2000 where he is currently an associate professor of physics and biophysics. His honors include 2001 Searle scholar, 2002 NSF CAREER award, 2003 Sloan fellowship, and 2003 Cottrell scholar.

## Acknowledgements
We thank members of Ha laboratory for their contributions, and long time collaborators David Lilley and Timothy Lohman for their continued support. Funding was provided by the US National Science Foundation (PHY-0134916, DBI-0215869) and by the US National Institutes of Health (GM065367, T.H.). S.A.M. was supported by the NIH molecular biophysics training grant (T32GM008276) and by a NSF graduate fellowship.